\documentclass[fleqn,12pt,twoside]{article}
\usepackage[headings]{espcrc1}

\readRCS
$Id: espcrc1.tex,v 1.2 2004/02/24 11:22:11 spepping Exp $
\usepackage{graphicx}
\usepackage[figuresright]{rotating}

\newcommand{\AmS}{{\protect\the\textfont2
  A\kern-.1667em\lower.5ex\hbox{M}\kern-.125emS}}

\title{Pygmy resonances probed with electron scattering}

\author{C. A. Bertulani
\address[MCSD]{
Department of Physics and Astronomy, University of Tennessee
Knoxville, TN 37996-1200 and
Physics Division, Oak Ridge National Laboratory, Oak Ridge, TN 37831
}%
        \thanks{Supported by the U.\thinspace S.\
Department of Energy under grants DE-FG02-04ER41338 and SciDAC-UNEDF.}}

\runtitle{Pygmy resonances probed with electron scattering}
\runauthor{C.A. Bertulani}

\begin{document}

% typeset front matter
\maketitle

\begin{abstract}
Pygmy resonances in light nuclei excited in
electron scattering are discussed. These collective
modes will be explored in future
electron-ion colliders
such as ELISe/FAIR (spokesperson: Haik Simon - GSI).
Response functions for direct breakup are explored
with few-body and hydrodynamical models, including
the dependence upon final state interactions.

\end{abstract}

\section{Introduction}

Reactions with radioactive beams have attracted great experimental
and theoretical interest  during the last two decades \cite{BHM02}.
Progresses of this scientific endeavor were reported on
measurements of nuclear sizes \cite{Ta85}, use of secondary
radioactive beams to elucidate reactions of
astrophysical interest \cite{BBR86,BB88}, fusion reactions with
neutron-rich nuclei \cite{TS91,Hus92}, tests of fundamental
interactions \cite{Har97}, dependence of the equation of state of
nuclear matter upon the asymmetry energy \cite{DLL02}, and many other
research topics.

New research areas with nuclei far from the stability line will become possible with
newly proposed experimental facilities. One of the projects
at the future FAIR facility of the GSI laboratory/Germany is
the study of electron scattering off unstable nuclei in an electron-ion
collider mode \cite{Haik05}. A similar proposal exists for the
RIKEN\ laboratory facility in Japan \cite{Sud01}. By means of elastic electron
scattering, these facilities will become the main probe of charge
distribution in unstable nuclei \cite{Ant05,Ber06}.
Coulomb excitation
has been very useful in determining the electromagnetic
response in light nuclei  \cite{BCH93}. For neutron-rich isotopes
\cite{Lei01} the resulting photo-neutron cross sections are
characterized by a pronounced concentration of low-lying $E1$
strength.  But it is
well known that non-perturbative effects leading to distortion of the energy
spectrum of the fragments interacting with the target (see, e.g. ref.
\cite{Ber05}) is a problem of difficult nature. The nuclear
response probed with electrons is free from such effects.

The interpretation of the low-lying $E1$ strength in neutron-rich nuclei
engendered a debate:
are these \textquotedblleft soft dipole modes\textquotedblright\
just a manifestation of the loosely-bound character of light
neutron-rich nuclei, or are they the result of the excitation of a resonance?
\cite{Iek93,Sac93,Sag95,Hus96}.  The
electromagnetic response in light nuclei, leading to their
dissociation, is related to the nuclear physics
needed in several astrophysical sites \cite{BBR86,BB88,Ber05}.  The existence of
pygmy resonances have important implications on theoretical
predictions of radiative neutron capture rates in the r-process
nucleosynthesis, and consequently on the calculated elemental
abundance distribution in the universe \cite{Go98}.

\section{Inelastic scattering in electron-ion colliders}

Here, $J_{i}$ $\left(  J_{f}\right)  $ $\ $is the initial (final)
angular momentum of the nucleus, $\left(  E,\mathbf{p}\right)  $ and
($E^\prime ,\mathbf{p}^{\prime}$) are the initial and final energy
and momentum of the electron, and $\left(  q_{0},\mathbf{q}\right)
=\left(  (E-E^{\prime})/\hbar c,\left(  \mathbf{p-p}^{\prime}\right)
/\hbar\right)  $ is the energy and momentum transfer in the
reaction.
For low energy excitations, $E,E^{\prime}\gg\hbar cq_{0}$,
for electron energies \ $E\simeq500$ MeV and
excitation energies $\Delta E=\hbar cq_{0}\simeq1-10$ MeV.

In the plane wave Born approximation (PWBA) and using the Siegert's
theorem, one can show that \cite{Ber062}
\begin{equation}
\frac{d\sigma}{d\Omega dE_{\gamma}}=\sum_{L}\frac{dN_{e}^{(EL)}\left(
E,E_{\gamma},\theta\right)  }{d\Omega dE_{\gamma}}\ \sigma_{\gamma}%
^{(EL)}\left(  E_{\gamma}\right)  ,\label{EPA}%
\end{equation}
where $\sigma_{\gamma}^{(EL)}\left(  E_{\gamma}\right)  \propto dB\left(
EL\right)  /dE_{\gamma}$,  is the photo-nuclear cross section for the $EL$%
-multipolarity, and  $E_{\gamma
}=\hbar cq_{0}$.
The response function, $dB\left(
EL\right)  /dE_{\gamma},$  is given by
\begin{equation}
\frac{dB\left(  EL\right)  }{dE_{\gamma}}=\frac{\left\vert \left\langle
J_{f}\left\Vert Y_{L}\left(  \widehat{\mathbf{r}}\right)  \right\Vert
J_{i}\right\rangle \right\vert ^{2}}{2J_{i}+1}\left[  \int_{0}^{\infty
}dr\ r^{2+L}\ \ \delta\rho_{if}^{\left(  EL\right)  }\left(  r\right)
\right]  ^{2}\rho\left(  E_{\gamma}\right)  ,\label{photo}%
\end{equation}
where $\rho\left(  E_{\gamma}\right)  $ is the density of final states
(for nuclear excitations into the continuum) with energy $E_{\gamma}%
=E_{f}-E_{i}$. The geometric coefficient $\left\langle J_{f}\left\Vert
Y_{L}\left(  \widehat{\mathbf{r}}\right)  \right\Vert J_{i}\right\rangle $ and
the transition density $\delta\rho_{if}^{\left(  EL\right)  }\left(  r\right)
$ will depend upon the nuclear model adopted.

One can also define a differential cross section integrated over angles so
that
\begin{equation}
\frac{dN_{e}^{(EL)}\left(  E,E_{\gamma}\right)  }{dE_{\gamma}}=2\pi
\int_{E_{\gamma}/E}^{\theta_{m}}d\theta\ \sin\theta\ \frac{dN_{e}%
^{(EL)}\left(  E,E_{\gamma},\theta\right)  }{d\Omega dE_{\gamma}},\label{EPA3}%
\end{equation}
and $\theta_{m}$ is the maximum electron scattering angle, which depends upon
the experimental setup. \ Notice that the lowest limit in the above integral
is $\theta_{\min}=E_{\gamma}/E$, and not zero.

Eqs. \ref{EPA}-\ref{EPA3} show that under the conditions of the
proposed electron-ion colliders, electron scattering will offer the
same information obtained with real photons. The
reaction dynamics information is in the virtual photon
spectrum, $N_{e}^{(EL)}\left(
E,E_{\gamma},\theta\right)$, while the nuclear response
dynamics information is in eq. \ref{photo}.
$dN_{e}^{(EL)}/d\Omega dE_{\gamma}$ is interpreted as
the number of equivalent (real) photons incident on the nucleus per
unit scattering angle $\Omega$ and per unit photon energy
$E_{\gamma}.$ For larger
scattering angles the Coulomb multipole matrix elements
are in general larger than the electric ($EL$)
multipoles, and monopole transitions become relevant \cite{Sch54}.
Eq. \ref{EPA} will not be valid under these
conditions.

It is found that the spectrum ${dN_{e}^{(EL)}\left(
E,E_{\gamma}\right)  }/{dE_{\gamma}}$ increases rapidly with
decreasing energies  \cite{Ber062}. For $E=500$ MeV and
excitation energies $\Delta E= 1$ MeV, the spectrum yields the ratios $dN_{e}%
^{(E2)}/dN_{e}^{(E1)}\simeq500$ and
$dN_{e}^{(E3)}/dN_{e}^{(E2)}\simeq100$. However, although
$dN_{e}^{(EL)}/dE_{\gamma}$ increases with the multipolarity $L$,
the nuclear response decreases rapidly with $L$, and $E1$
excitations tend to dominate the cross sections. For larger electron
energies the ratios $N^{(E2)}/N^{(E1)}$ and $N^{(E3)}/N^{(E1)}$
decrease rapidly. A comparison between the $E1$ virtual photon
spectrum, $dN_{e}^{(E1)}/dE_{\gamma}$, of 1 GeV electrons with the spectrum
generated by 1 GeV/nucleon heavy ion projectiles was published in ref.
\cite{Ber062}. The virtual spectrum for the electron contains more
hard photons, i.e. the spectrum decreases slower with photon energy
than the heavy ion photon spectrum. This is because, in both
situations, the rate at which the spectrum decreases depends on the
ratio of the projectile kinetic energy to its rest mass, $E/mc^{2},$
which is much larger for the electron ($m=m_{e}$) than for the heavy
ion ($m=$ nuclear mass).

\section{Dissociation of weakly-bound systems}

\subsection{One-neutron halo}

In a two-body model, the single-particle picture has been used previously to study Coulomb
excitation of halo nuclei with success \cite{BB86,BS92,Ots94,MOI95,KB96,TB04}.
The $E1$
transition integral $\mathcal{I}_{l_{i}l_{f}}=\int_{0}^{\infty}dr\
r^3\ \ \delta\rho_{if}\left(
r\right)  $ becomes
\begin{equation}
\mathcal{I}_{s\rightarrow p}
  \simeq\frac{e_{ff}\hbar^{2}}{2\mu}\frac{2E_{r}}{\left(  S_{n}+E_{r}\right)
^{2}}\left[  1+\left(  \frac{\mu}{2\hbar^{2}}\right)  ^{3/2}\frac{\sqrt{S_{n}%
}\left(  S_{n}+3E_{r}\right)  }{-1/a_{1}+\mu r_{1}E_{r}/\hbar^{2}}\right]
,\label{isp}%
\end{equation}
in terms of the effective range expansion of the phase shift, $k^{2l+1}\cot
\delta\simeq-1/a_{l}+r_{l}k^{2}/2$.

As shown in previous works \cite{BB86,BS92}, a peak
is manifest in the response function, $
{dB(EL)}/{dE}\propto\left\vert \mathcal{I}_{s\rightarrow p}\right\vert
^{2}\propto{E_{r}^{L+1/2}}/{\left(  S_{n}+E_{r}\right)  ^{2L+2}%
}.$ It appears centered at the energy \cite{BS92} $
E_{0}^{(EL)}\simeq({L+1/2})S_{n}/({L+3/2})$
for a generic electric response of multipolarity $L$. For $E1$ excitations,
the peak occurs at $E_{0}\simeq3S_{n}/5$.
The second term inside brackets in eq. \ref{isp} is a modification
due to final state interactions. It is important in many cases \cite{Ber062}.

\subsection{Two-neutron halo}

Many weakly-bound nuclei, like $^{6}$He or $^{11}$Li, require a three--body
treatment. In one particular model, the bound--state wavefunction in the center of mass system is
written as an expansion over hyperspherical harmonics, see
e.g.~\cite{Zhu93}.
For weakly-bound systems having
no bound subsystems the hyperradial functions behave asymptotically as \cite{Mer74}
$\Phi_{a}\left(  \rho\right)  \longrightarrow \exp\left(
-\eta\rho\right)$ as $\rho\longrightarrow
\infty$,
where the two-nucleon separation energy is related to $\eta$ by $S_{2n}%
=\hbar^{2}\eta^{2}/\left(  2m_{N}\right)  $.
(see also
\cite{Chu93}). The $E1$ transition matrix element is obtained by a
sandwich of the E1 operator between $\Phi_{a}\left(  \rho\right)
/\rho^{5/2}$ and scattering wavefunctions. Following ref. \cite{Pus96}, but
using distorted
scattering states,
\begin{equation}
\mathcal{I}\left(  E1\right)  =\int dxdy\frac{\Phi_{a}\left(  \rho\right)
}{\rho^{5/2}}\ y^{2}x u_{p}\left(  y\right)   u_{q}\left(  x\right)
,\label{ie1}%
\end{equation}
where $u_{p}\left(  y\right)  =j_{1}\left(  py\right)  \cos\delta_{nc}%
-n_{1}\left(  py\right)  \sin\delta_{nc}$ is the core-neutron asymptotic
continuum wavefunction, assumed to be a $p$-wave, and $u_{q}\left(  x\right)
=j_{0}\left(  qx\right)  \cos\delta_{nn}-n_{0}\left(  qx\right)  \sin
\delta_{nn}$ is the neutron-neutron asymptotic continuum wavefunction, assumed
to be an $s$-wave.

The $E1$ strength function is proportional to the square of
eq. \ref{ie1} integrated over all momentum variables.
As pointed out in ref. \cite{Pus96}, the $E1$ three-body response function of
$^{11}$Li can still be described by an expression similar to the two-body case, but with
different factors. Explicitly, $
{dB\left(  E1\right)  }/{dE_{r}}\propto{E_{r}^{3}}/{\left(
S_{2n}^{eff}+E_{r}\right)  ^{11/2}}.$
Instead of $S_{2n}$, one has to use an effective $S_{2n}^{eff}=aS_{2n}$,
with $a\simeq1.5$. With this approximation, the peak of the strength function
in the three-body case is situated at about
three times higher energy than for the two-body case. In the
three-body model, the maximum is thus predicted at $E_{0}^{(E1)}%
\simeq1.8S_{2n}$, which fits the experimental peak position for
the $^{11}$Li $E1$ strength function very well \cite{Pus96}. It is thus
apparent that the effect of three-body configurations is to widen and to shift
the strength function $dB\left(  E1\right)  /dE$ to higher energies.

\subsection{The hydrodynamical model}

As with giant dipole resonances (GDR) in stable
nuclei, one believes that pygmy resonances at energies close to
threshold are present in halo, or neutron-rich, nuclei. This was
proposed by Suzuki et al. \cite{SIS90} using the hydrodynamical
model for collective vibrations.
We will use the method of Myers et al.
\cite{Mye77}, considering collective vibrations in nuclei as
an admixture of Goldhaber-Teller (GT) and Steinwedel-Jensen (SJ) modes. For light
nuclei Goldhaber-Teller modes dominate.

For spherically symmetric densities, the transition density,
$\delta\rho_{p}\left(  {\bf r}\right)=\delta\rho_{p}
\left(  r\right)  Y_{10}\left(  \widehat{\mathbf{r}}\right)$, can
be calculated assuming a  combination of SJ and GT
distributions \cite{Ber062},
\begin{equation}
\delta\rho_{p}\left(  r\right)  =\sqrt{\frac{4\pi}{3}}R\left\{  Z_{eff}%
^{(1)}\alpha_{1}\frac{d}{dr}+Z_{eff}^{(2)}\alpha_{2}\frac{K}{R}j_{1}\left(
kr\right)  \right\}  \rho_{0}(r),\label{transdsjgt}%
\end{equation}
where
$R$ is the mean nuclear radius, and $\alpha_{i}$
is the percent displacement of the center of mass of
neutron and proton fluids in the GT ($i=1$) and SJ ($i=2$) modes.
For light, weakly-bound nuclei, it is appropriate to assume
that the neutrons inside the core ($A_{c},Z_{c}$) vibrate in phase
with the protons. The neutrons and protons in the core are tightly
bound.
Calling the excess nucleons by
$(A_{e},Z_{e})=(A-A_{c},Z-Z_{c}),$ the
effective charge for the GT mode is $Z_{eff}%
^{(1)}=$ $\left(  Z_{c}A_e  -A_{c}Z_e
\right)  /A$.
In eq. \ref{transdsjgt}, $j_1(kr)$ is the spherical Bessel function of first order,
$\alpha_{2}$ the percent displacement of the center of mass of the
neutron and proton
fluids in the SJ mode,
$kR=a=2.081,\ \ \ \ {\rm and}\ \ K={2a}/{j_{0}\left(  a\right)
}=9.93.$

The hydrodynamical model can be further explored to obtain the energy and
excitation strength of the collective excitations. This can be achieved by finding
the eigenvalues of the Hamiltonian $
\mathcal{H}=\frac{1}{2}\dot{\alpha}\mathcal{T}\dot{\alpha}+\frac{1}{2}%
\alpha\mathcal{V}\alpha+\dot{\alpha}\mathcal{F}\dot{\alpha},\label{ldh}%
$
where $\alpha=\left(  \alpha_{1},{\rm  }\alpha_{2}\right)  $ is
now a vector containing GT and SJ contributions to the
collective motion. $\mathcal{T}$ and $\mathcal{V}$ are the kinetic and potential
energies $2\times2$ matrices \cite{Mye77}.
The kinetic term  can be calculated from GT and SJ velocity fields.
The potential term  is related to the stiffness
parameters of the liquid-drop model and adjusted to a best fit to the
nuclear masses. The stiffness of the system is due to the change in
symmetry energy of the system
as it goes out of the equilibrium position. The last term in the Hamiltonian
is the Rayleigh dissipation term, which is related to
the Fermi velocity of the nucleons \cite{Mye77} and yields the width of
the eigenstate.

As shown by Myers et al. \cite{Mye77}, the liquid drop model
predicts an equal admixture of SJ+GT oscillation modes for large
nuclei. The contribution of the SJ oscillation mode decreases with
decreasing mass number, i.e. $\alpha\longrightarrow\left(
\alpha_{1},{\rm \ }0\right)  $ as $A\longrightarrow0$. This is even
more probable in the case of halo nuclei, where a special type of GT
mode (oscillations of the core against the halo nucleons) is likely
to be dominant. For this special collective motion an approach
different than those used in refs. \cite{Mye77} and \cite{SIS90} has
to be considered.

It is easy to make changes in the original Goldhaber and Teller \ \cite{GT48}
formula to obtain the energy of the collective vibrations, yielding
\begin{equation}
E_{PR}=\left(  \frac{3\varphi\hbar^{2}}{2aRm_{N}A_{r}}\right)  ^{1/2}%
,\label{EPGDR}%
\end{equation}
where $A_{r}=A_{c}\left(  A-A_{c}\right)  /A$ and $a$ is the length
within which the interaction between a neutron and a nucleus changes
from a zero-value outside the nucleus to a high value inside, i.e.
$a$ is the width of the nuclear surface. $\varphi$ is the energy
needed to extract one neutron from the proton environment. Goldhaber
and Teller \cite{GT48} argued that in a heavy stable nucleus
$\varphi$ is not the binding energy of the nucleus, but the part of
the potential energy due neutron-proton asymmetry. In the case of
weakly-bound nuclei this picture changes and it is more reasonable
to associate $\varphi$ to the separation energy of the valence
neutrons, $S$. I will use $\varphi=\beta S$, with a parameter
$\beta$ which is expected to be of order of one. Since for halo
nuclei the product $aR$ \ is proportional to $S^{-1},$ we obtain the
proportionality $E_{PR}\propto S$.  Using eq. \ref{EPGDR} for
$^{11}$Li , with \ $a=1$ fm, $R=3$ fm and $\varphi=S_{2n}=0.3$ MeV,
we get $E_{PR}=1.3$ MeV. Considering that the pygmy resonance will
most probably decay by particle emission, one gets $E_{r}\simeq1$
MeV for the kinetic energy of the fragments. This is about a factor
2 larger than what is obtained in a numerical calculation
\cite{Ber062}. But it is within the right ballpark. The
hydrodynamical model is very unlikely to be an accurate model for
light, loosely-bound, nuclei and is significant only in that a
reasonable magnitude of the resonance energy is found.

Both the direct dissociation model and the hydrodynamical model
yield a bump in the response function with position proportional to
$S$, the valence nucleon(s) separation energy.  In the direct
dissociation model the width of the response function depends on the
separation energy and on the nature of the model, i.e. a
two- or a three-body model. In the case of the pygmy resonance, this
question is completely open. The hydrodynamical model predicts
\cite{Mye77} for the width of the collective mode $\Gamma
=\hbar\overline{\mathrm{v}}/R$, where $\overline{\mathrm{v}}$ \ is
the average velocity of the nucleons inside the nucleus. This
relation can be derived by assuming that the collective vibration is
damped by the incoherent collisions of the nucleons with the walls
of the nuclear potential well during the vibration cycles. This
approach mimics that used in the kinetic theory of gases for
calculating the energy transfer of a moving piston to gas molecules
in a container. Using $\overline{\mathrm{v}}=3\mathrm{v}_{F}/4$,
where $\mathrm{v}_{F}=\sqrt {2E_{F}/m_{N}}$ is the Fermi velocity,
with $E_{F}=35$ MeV and $R=6$ fm, one gets $\Gamma\simeq6$ MeV. This
is the typical energy width a giant dipole resonance state in a
heavy nucleus. But in the case of neutron-rich light nuclei
$\overline{\mathrm{v}}$ is not well defined. But the piston model is not
able to describe the width of the response function properly.
Microscopic models, such as those based on random phase
approximation (RPA) calculations, are necessary to tackle this
problem. The halo nucleons have to be treated in an special way to
get the response at the right energy position, and with
approximately the right width. Right now, the problem remains if the
experimentally observed peak in $dB/dE$ is due to a direct
transition to the continuum, weighted by the phase space of the
fragments, or if it proceeds sequentially via a soft dipole
collective state.


\begin{thebibliography}{9}

\bibitem {BHM02}C.A. Bertulani, M.S. Hussein, and G. M\"{u}nzenberg,
\textit{Physics of Radioactive Beams} (Nova Science Publishers, Hauppage, NY, 2002).

\bibitem {Ta85}I. Tanihata et al., Phys. Rev. Lett. 55 (1985) 2676.

\bibitem {BBR86}G.Baur, C.A.Bertulani and H.Rebel, Nucl. Phys. A 458
(1986) 188.

\bibitem {BB88}C.A.Bertulani and G.Baur, Phys. Reports 163 (1988) 299.

\bibitem {TS91}N. Takigawa and H. Sagawa, Phys. Lett. B 265 (1991) 23.

\bibitem {Hus92}M. S. Hussein et al., Phys. Rev. C 46 (1992) 377.

\bibitem {Har97}J. Hardy, in \textit{Physics of Unstable Nuclear Beams},
Edited by C.A.Bertulani et al., World Scientific, Singapore, 1997.

\bibitem {DLL02}P. Danielewicz, R. Lacey and W. G. Lynch, Science
298, 1592 (2002).

\bibitem {Haik05}Haik Simon, \textit{Technical Proposal for the Design,
Construction, Commissioning, and Operation of the ELISe setup}, GSI Internal
Report, Dec. 2005.

\bibitem {Sud01}T. Suda, K. Maruayama, \textit{Proposal for the RIKEN beam
factory}, RIKEN, 2001; M. Wakasugia, T. Suda, Y. Yano, Nucl. Inst.
Meth. Phys. A 532, 216 (2004).

\bibitem {Ant05}A.N. Antonov et al., Phys. Rev. C 72, 044307 (2005).

\bibitem {Ber06}C.A. Bertulani, J. Phys. G 34 (2007) 315.

\bibitem {BCH93}C.A. Bertulani, L.F. Canto and M.S. Hussein, Phys. Reports
226(1993) 281.

\bibitem {Lei01}A. Leistenschneider et al., Phys. Rev. Lett. 86, 5442 (2001).

\bibitem {Ber05}C.A.Bertulani, Phys. Rev. Lett. 94, 072701 (2005).

\bibitem {Iek93}K. Ieki et al., Phys. Rev. Lett. 70, 730 (1993).

\bibitem {Sac93}D. Sackett et al., Phys. Rev. C 48, 118 (1993).

\bibitem {Sag95}H. Sagawa et al., Z. Phys. A 351, 385 (1995).

\bibitem {Hus96}M.S. Hussein, C.Y Lin and A.F.R. de Toledo Piza, Z. Phys.
A 355 (1966) 165.

\bibitem {Go98} S. Goriely, Phys. Lett. B 436 (1998) 10.

\bibitem {Ber062} C.A. Bertulani, Phys. Rev.  C, in press.

\bibitem {Sch54}L.I. Schiff, Phys. Rev. 96, 765 (1954).

\bibitem {BB86}C.A. Bertulani and G. Baur, Nucl. Phys. A 480, 615
(1988).

\bibitem {BS92}C.A. Bertulani and A. Sustich, Phys. Rev. C 46, 2340 (1992).

\bibitem {Ots94}T. Otsuka et al., Phys. Rev. C 49, R2289 (1994).

\bibitem {MOI95}A. Mengoni, T. Otsuka and M. Ishihara, Phys. Rev. C
52, R2334 (1995).

\bibitem {KB96}D.M. Kalassa and G. Baur, J. Phys. G 22, 115 (1996).

\bibitem {TB04}S. Typel and G. Baur, Phys. Rev. Lett. 93, 142502 (2004).

\bibitem {Zhu93}M.~V. Zhukov, B.V. Danilin, D.V. Fedorov, J.M. Bang, I.J.
Thompson and J.S. Vaagen, Phys. Rep. 231, 151 (1993).

\bibitem {Mer74}S.P. Merkuriev, Sov. J. Nucl. Phys. 19 (1974) 222.

\bibitem {Pus96}A Pushkin, B Jonson and M V Zhukov, J. Phys. G 22
(1996) L95.

\bibitem {Chu93}L.V. Chulkov, B. Jonson and M.V. Zhukov, J. Phys. G
22 (1996) 95.

\bibitem {SIS90}Y. Suzuki, K. Ikeda, and H. Sato, Prog. Theor. Phys.
83, 180 (1990).

\bibitem {GT48}M. Goldhaber and E. Teller, Phys. Rev. 74, 1046 (1948).

\bibitem {SJ50}H. Steinwedel and H. Jensen, Z. Naturforschung 5A, 413 (1950).

\bibitem {Mye77}W.D. Myers, W.G. Swiatecki, T. Kodama, L.J. El-Jaick and E.R.
Hilf, Phys. Rev. C 15, 2032 (1977).


\end{thebibliography}
\end{document}